\documentclass[aip,graphicx]{revtex4-1}

\usepackage{graphicx}
\usepackage{url}
\usepackage{color}

\newcommand*{\citen}[1]{%
  \begingroup
    \romannumeral-`\x % remove space at the beginning of \setcitestyle
    \setcitestyle{numbers}%
    \cite{#1}%
  \endgroup   
}

\draft % marks overfull lines with a black rule on the right

\begin{document}

%\preprint{}

\title{Use of Cernox thermometers in AC specific heat measurements under pressure} %Title of paper



\author{Elena Gati}
%\thanks{}
\affiliation{Ames Laboratory, US Department of Energy, Iowa State University, Ames, Iowa 50011, USA}
\affiliation{Department of Physics and Astronomy, Iowa State University, Ames, Iowa 50011, USA}

\author{Gil Drachuck}
\affiliation{Ames Laboratory, US Department of Energy, Iowa State University, Ames, Iowa 50011, USA}
\affiliation{Department of Physics and Astronomy, Iowa State University, Ames, Iowa 50011, USA}

\author{Li Xiang}
\affiliation{Ames Laboratory, US Department of Energy, Iowa State University, Ames, Iowa 50011, USA}
\affiliation{Department of Physics and Astronomy, Iowa State University, Ames, Iowa 50011, USA}

\author{Lin-Lin Wang}
\affiliation{Ames Laboratory, US Department of Energy, Iowa State University, Ames, Iowa 50011, USA}

\author{Sergey L. Bud'ko}
\affiliation{Ames Laboratory, US Department of Energy, Iowa State University, Ames, Iowa 50011, USA}
\affiliation{Department of Physics and Astronomy, Iowa State University, Ames, Iowa 50011, USA}

\author{Paul C. Canfield}
\affiliation{Ames Laboratory, US Department of Energy, Iowa State University, Ames, Iowa 50011, USA}
\affiliation{Department of Physics and Astronomy, Iowa State University, Ames, Iowa 50011, USA}

\date{\today}

\begin{abstract}
We report on the resistance behavior of bare-chip Cernox thermometers under pressures up to 2\,GPa, generated in a piston-cylinder pressure cell. Our results clearly show that Cernox thermometers, frequently used in low-temperature experiments due to their high sensitivity, remain highly sensitive even under applied pressure. We show that these thermometers are therefore ideally suited for measurements of heat capacity under pressure utilizing an ac oscillation technique up to at least 150\,K. Our Cernox-based system is very accurate in determining changes of the specific heat as a function of pressure as demonstrated by measurements of the heat capacity on three different test cases: (i) the superconducting transition in elemental Pb ($T_c\,=\,7.2\,$K), (ii) the antiferromagnetic transition in the rare-earth compound GdNiGe$_3$ ($T_N\,=\,26\,$K) and (iii) the structural/magnetic transition in the iron-pnictide BaFe$_2$As$_2$ ($T_{s,N}\,=\,130\,$K). The chosen examples demonstrate the versatility of our technique for measuring the specific heat under pressure of various condensed-matter systems with very different transition temperatures as well as amounts of removed entropy. 
\end{abstract}


\maketitle %\maketitle must follow title, authors, abstract and \pacs

\section{Introduction}

The specific heat of solids is one of the most fundamental thermodynamic quantities. Its temperature dependence reveals important information about the energy scales of electronic, magnetic and lattice degrees of freedom. It is thus an inherently sensitive tool to detect phase transitions which involve one or more of the above-mentioned degrees of freedom. Consequently, various techniques \cite{Stewart83} to measure specific heat in calorimetric experiments at ambient pressure and low temperatures are well established and nowadays even commercially available \cite{Lashley03,acoption}. Typically, these techniques require adiabatic conditions, i.e., an almost ideal decoupling of the sample from the environment, so as to achieve high accuracy in the determination of absolute values of the specific heat. Among these techniques, the relaxation method \cite{Bachmann72} is considered as the standard method in which a heat pulse is given to the sample of interest and the relaxation time towards the initial temperature after switching off the pulse is directly related to the size of the specific heat. Alternatively, in particular in cases where the sample mass is very small, is the AC technique \cite{Sullivan68,Eichler79,Kraftmakher02}, in which the sample is heated by an oscillatory heat source and the resulting temperature oscillation can be used to infer the specific heat, preferred.

 As a matter of fact, AC specific heat measurements have proven to be particularly suited for measurements under pressure \cite{Bonilla74,Baloga77,Eichler79,Chen93,Bouquet00,Demuer00,Wilhelm03,Kubota08,Umeo17}. In general, pressure $p$ represents an essential parameter for tuning phase transitions in solids \cite{Jayaraman72,Lorenz05,Imada98,Brando16}. To perform measurements under pressure, the sample has to be embedded into a pressure medium inside a pressure cell. This typically provides a stronger coupling between the sample and the bath (i.e., the oustide of the pressure cell), compared to ambient-pressure experiments performed in vacuum. Whereas the analysis of data taken under pressure with the relaxation method suffers from the huge addenda contribution from pressure cell and medium, the AC technique has a second major advantage in addition to its sensitivity for samples with small masses: the choice of the measurement frequency allows for the measurement on a different timescale than the one determined by the relaxation time to the bath. This can result, to first approximation, in a decoupling of the sample from the bath, thereby paving the way to extraction of absolute values of the specific heat on a semi-quantitative level \cite{Eichler79}.
 
 Typically, in order to perform AC calorimetric experiments in piston-pressure cells up to $p\,\approx\,2\,-\,3\,$GPa, either small ruthenium oxide (RuO$_2$) thermometers \cite{Kubota08,Chen93,Baloga77} or thermocouples \cite{Bonilla74,Wilhelm03,Bouquet00,Demuer00} have been used to detect the temperature oscillations. On one hand, RuO$_2$ thermometers are inherently sensitive only at low temperatures due to their insulating nature; on the other hand they are easy to handle. Thermocouples cover a wider temperature range, but come along with obstacles in their handling. The reliable use of thermocouples requires a firm contact to the sample which often can be only realized by spot-welding of the thermocouple to the sample. Spot-welding is not possible in case of non-metallic samples, but also is often found to be problematic for metallic (and often brittle) samples. In addition, obtaining absolute values of temperature changes with high accuracy can only be guaranteed when the thermal contact to a reference temperature is good which can be challenging in the pressure-cell environment. 
 
 In this work, we present another option: using Cernox\cite{Cernox} thermometers as temperature sensors. Cernox sensors combine the advantages of RuO$_2$ thermometers and thermocouples. They are well established at ambient pressure in most low-temperature laboratories, as they provide a high sensitivity over a wide temperature range, as well as short thermal response times. In addition, they can easily be attached to any sample without the need of spot-welding. This being said, it is surprising that, to the best of our knowledge, the properties of Cernox thermometers have not been studied under pressure so far. Our results show that the sensitivity of Cernox thermometers remains large over the entire investigated pressure range up to 2\,GPa and temperature range up to at least 150\,K. We demonstrate that this high sensitivity of the sensors allows us to study the specific heat of solids under pressure (including various types of phase transitions) at a semi-quantitative level. The wide temperature range covered by this setup will allow for the study of larger regions of phase diagrams by specific heat under pressure, with the convenience of using commercially-available temperature sensors.
 
 This paper is organized as follows. First, we describe details of the experimental setup (Sec. \ref{sec:setup}) used in this work to determine the resistance behavior of the Cernox thermometers under pressure, as well as the specific heat of solids under pressure. In the next section (Sec. \ref{sec:Cernoxresistance}), we show one of the main results of this work, namely that the resistance change of the Cernox thermometers under pressures up to 2\,GPa is very modest and readily describable. Following this, we turn to our description of the AC specific heat data obtained using these Cernox thermometers. Therefore we first provide some theoretical background information on AC specific heat measurements and illustrate our measurement protocol in Sec. \ref{sec:ACtheory}. In Sec. \ref{sec:specheatresults} we discuss the results of specific heat under pressure measurements on three different test cases each of which undergoes a different type of phase transition. These systems were chosen to cover a wide range of phase transition temperatures (7 K up to 130 K) as well as removed entropies, thereby demonstrating the versatility of Cernox thermometers for measurements of specific heat under pressure.
 
\section{Experimental Setup}
\label{sec:setup}

\begin{figure}
				\includegraphics[width=0.9\textwidth]{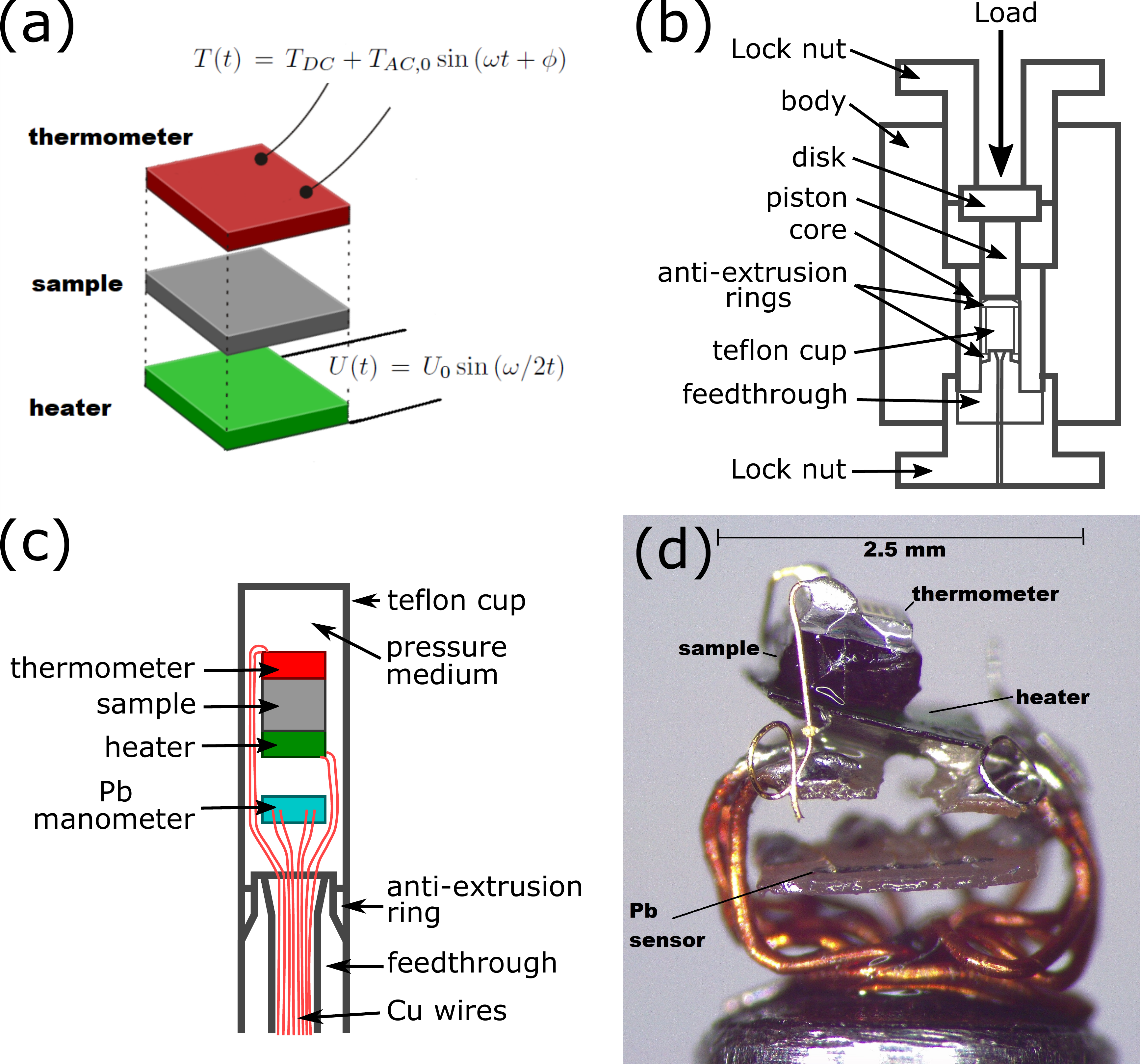} 
				\caption{(a) Schematic diagram of the sample arrangement with heater and thermometer. The heater is supplied with an AC voltage with frequency $\omega/2$ which results in an oscillation of the temperature of the sample with frequency $\omega$; (b) Schematic diagram of the piston-pressure cell used in the present work; (c) Schematic diagram of the sample assembly and Pb sensor inside the teflon cup; (d) Photograph of the heater, sample and thermometer mounted on the pressure-cell feedthrough. The Pb sensor for the determination of the pressure value near 7\,K is also mounted on the feedthrough.}
				\label{fig:schematicsetup}
				\end{figure}

To perform AC calorimetric measurements, the sample of interest is placed between a heater and a thermometer (see Fig.\,\ref{fig:schematicsetup} (a)). In our setup, we use bare Cernox-chip thermometers \cite{Cernox} (type CX-1070 or type CX-1080) as thermometers. The bare chips have dimensions of 0.965\,$\times\,0.762\,\times\,0.203$\,mm$^3$ and are thus ideally suited to fit into standard piston-pressure cells (see Figs.\,\ref{fig:schematicsetup} (b) and (c) for schematic drawings). In addition, they are deposited on a sapphire substrate with low mass (\,$\le\,3\,$mg), thus have themselves a small specific heat, and short response times (1.5\,ms at 4.2\,K). As a heater, we use strain gauges\cite{straingauges} (type FLG-02-23, Tokyo Sokki Kenkyujo Co., Ltd.) which have an active heater area of $\approx\,1\,\times\,1.4$\,mm$^2$. They show an almost temperature-independent resistance as a function of temperature ($R(T,p)\,\approx\,$120\,$\Omega$) and are enclosed in a very thin layer of plastic coating giving rise to a low thermal  mass. The samples, with typical masses $\,\sim\,2\,$mg, are cut into plates with dimensions as close as possible to the active heater area dimensions. The thermometer and heater are attached to the sample by using Devcon 5 Minute epoxy (No. 14250) to improve the thermal contact between the individual components and to guarantee sufficient mechanical stability in the pressure cell (shown schematically in Figs.\,\ref{fig:schematicsetup} (a) and (c)). A photograph of the assembly is shown in Fig.\,\ref{fig:schematicsetup}\,(d). The wires of the thermometer and heater are soldered to the wires passing the pressure-cell feedthrough. The thermometer is connected in a pseudo-four-point configuration in which the four wires for current and voltage are reduced to two wires inside the pressure cell. In addition, a Pb sample is mounted on the feedthrough in a four-point configuration for determining its critical temperature, $T_c$, via resistance measurements. The $T_c$ value can be used to determine the pressure, $p$, at low temperature as $T_c(p)$ is well characterized in literature \cite{Bireckoven88}. 

The sample end of the feedthrough is placed in a Teflon-cup (see Figs.\,\ref{fig:schematicsetup}\,(b) and (c)) which is filled with the pressure-transmitting medium. In all the experiments presented here, a mixture of 4:6 mixture of light mineral oil:n-pentane \cite{Budko84,Kim11} is used as a pressure-transmitting medium. It solidifies at $p\,\approx\,3-4\,$GPa at room temperature, thus ensuring hydrostatic pressure conditions in the available pressure range. Two anti-extrusion rings made out of phosphor-bronze are used to prevent the teflon from flowing through the interstices when pressurized. The outer cell body is made out of Grade 5 titanium alloy (Ti 6Al-4V) and the inner cylinder out of Ni-Cr-Al alloy. Its design is similar to the one described in Ref. \citen{Budko84}. As Ti 6Al-4V alloy turns superconducting \cite{Ridgeon17} below $\approx\,$5\,K and as a consequence, its thermal conductivity becomes significantly reduced, the sample inside the cell cannot be cooled below 5\,K. Therefore, the use of this particular cell is restricted to temperatures above 5\,K. This issue can be circumvented by using cells made out of a different material, such as CuBe/Ni-Cr-Al.

Pressure is applied by applying a load to the piston at room temperature by a hydraulic press and locked by tightening the lock  nut. All measurements shown in this manuscript were performed inside the pressure cell. At the beginning of each pressure cycle the pressure cell was closed hand-tight. Whereas this procedure typically results in a small, but finite pressure at room temperature \cite{Thompson84} ($p\,\lesssim\,0.3\,$GPa), the pressure at low temperature inferred from $T_c$ of Pb is usually very close to 0\,GPa ($p\,\lesssim\,0.04\,$GPa). We refer to this situation in the manuscript as ``ambient-pressure'' condition ($p\,=\,0\,$GPa). All data shown were obtained by increasing pressure to the measured value.

		The measurements were carried out in a cryogen-free cryostat (Janis SHI-950 with a base temperature of $\approx\,3.5\,$K). The probe, used in this cryostat, is wired with phosphor-bronze wires (QT-36, LakeShore Inc.) to ensure low heat flow through the wires. The temperature was controlled continuously between base and room temperature by a LakeShore 336 controller. Temperature was monitored by a calibrated temperature sensor (Cernox-1030) which was placed directly outside the pressure cell by inserting it into a copper bracket. The Cernox thermometer inside the pressure cell was supplied by a DC current (Model CS580, Stanford Research Systems). The size of the DC current was adjusted with temperature such that the voltage limit ($<\,100\,$mV) of the thermometer is not exceeded. The voltage oscillations of the thermometer which result from the AC heating were pre-amplified and filtered (Model SR560, Stanford Research Systems) and then measured with a Lock-In Amplifier (Model SR860, Stanford Research Systems) the internal oscillator of which was used to provide the heating voltage. The heating power was chosen such that the amplitude of the induced temperature oscillation $T_{AC,0}$ (see Fig.\,\ref{fig:schematicsetup} (a)) was typically smaller than 20\,mK. To measure frequency responses (i.e., measurements as a function of frequency, see below), we used the built-in frequency option of this particular Lock-In Amplifier which allows to change the frequency within user-defined frequency limits and sweep rates. The DC resistance of the bare Cernox chips inside the pressure cell was read out simultaneously to each specific heat measurement by a Digital Voltmeter (SIM970, Stanford Research Systems). The resistance of the Pb sensor was measured with a LakeShore AC Resistance Bridge (Model 370). All data are recorded using a custom LabView Program. 
		
		\section{Results: Cernox resistance under pressure}
		\label{sec:Cernoxresistance}
		
\begin{figure}
				\includegraphics[width=0.9\textwidth]{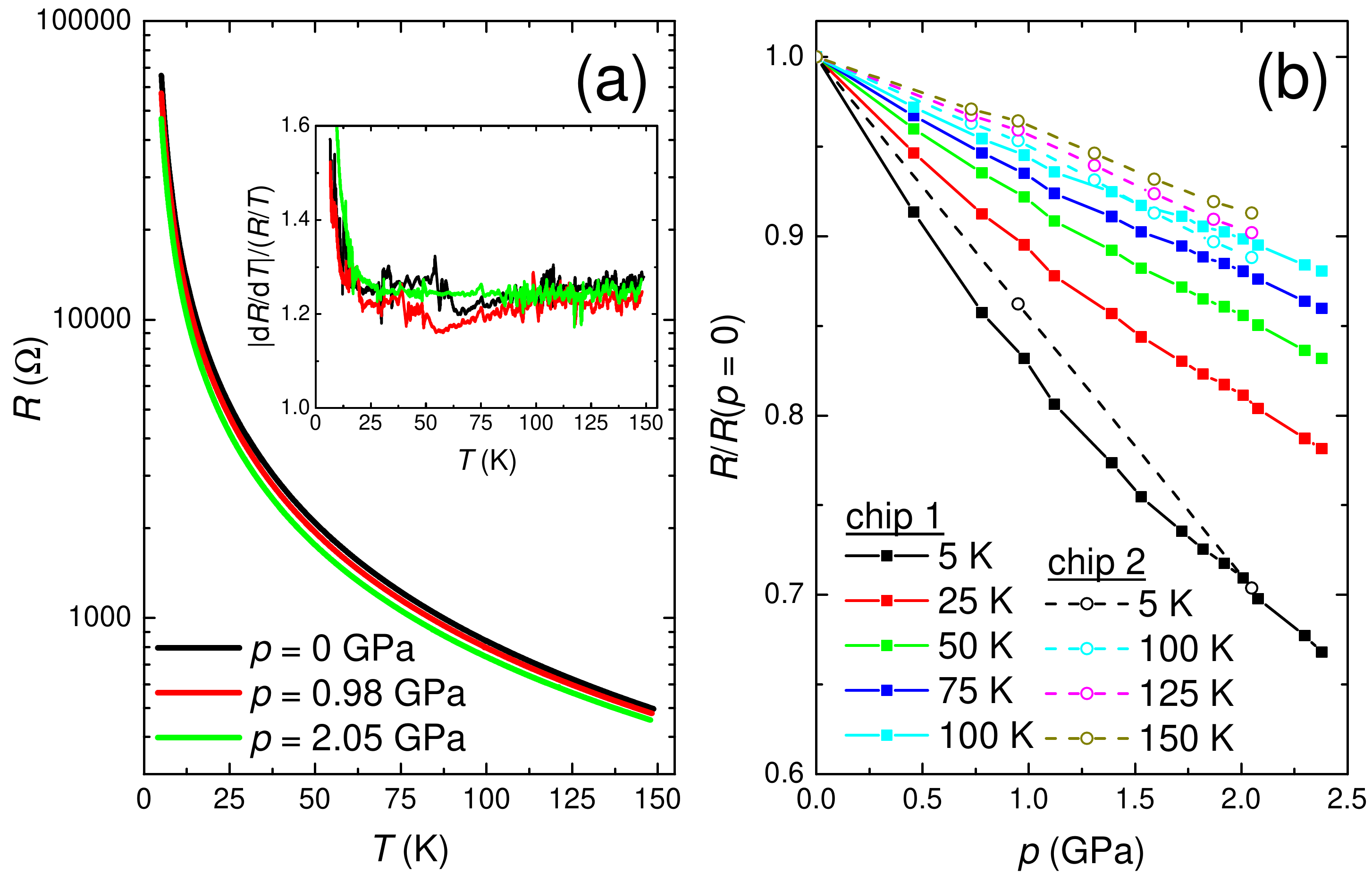} 
				\caption{(a) Resistance, $R$, of a Cernox thermometer, type CX-1080, as a function of temperature, $T$, at three different pressures $p$, ranging from 0\,GPa to 2.05\,GPa; Inset: Sensitivity, defined as $|$d$R/$d$T/(R/T)|$, as a function of temperature for the same pressure values, as depicted in the main panel. The step at $T\,\approx\,$50\,K in the data at 0\,GPa and 2.05\,GPa is likely an artifact associated with changing of thermometer current; (b) Resistance of a Cernox thermometer, type CX-1080, normalized to its ambient-pressure value, $R/R(p\,=\,0)$ at different temperatures between 5\,K and 150\,K. Open and closed symbols represent measurements on two different chips of the same type.}
				\label{fig:Cernoxresistance}
				\end{figure}

		Figure \ref{fig:Cernoxresistance} (a) summarizes our main result on the behavior of the Cernox (type CX-1080) resistance, $R$, as a function of temperature, $T$, at three selected pressures up to $p\,\approx\,$2\,GPa. These data were taken without any applied heat to the heater inside the pressure cell. At ambient pressure, the resistance shows a typical behavior for Cernox thermometers: the resistance increases with decreasing temperature and the slope d$R$/d$T$ is finite over the entire temperature range which guarantees a sufficient sensitivity of this thermometer from low temperatures ($T\,\approx\,$5\,K) up to high temperatures ($T\,\approx\,$150\,K). Upon increasing pressure, the resistance at a fixed temperature is reduced by $\approx\,$28\% at 5\,K ($\approx\,10\,$\% at 150\,K) at 2\,GPa (see Fig.\,\ref{fig:Cernoxresistance} (b) for change of resistance as a function of pressure at different temperatures). However, the overall behavior as a function of temperature is nearly unchanged. To quantify the sensitivity of the thermometer, one can refer to the dimensionless quantity (d$R$/d$T$)/($R/T$) which is displayed as a function of $T$ in the inset of Fig. \ref{fig:Cernoxresistance} (a) for the same pressure values as the ones depicted in the main panel. This representation shows that this type of Cernox thermometer has an almost temperature- and pressure-independent sensitivity factor of 1.25. Only at low temperatures ($T\,<\,25\,$K), an increased sensitivity up to 1.5 is observed for all pressures. Thus, our measurements clearly show that Cernox thermometers keep their high sensitivity across a wide temperature range up to 2\,GPa. Note that even though we restrict ourselves in this study to temperatures below 150\,K, it is known that (d$R$/d$T$)/($R/T$) of the Cernox thermometers remains almost unchanged at ambient pressure up to room temperature \cite{Cernox}. Based on our results, it is therefore likely that the Cernox thermometers are very sensitive up to room temperature, even under pressure. Moreover, we did not find any indications of changes in the thermometer behavior from one pressure cycle to the next or strong deviations in the behavior of different chips (see Fig.\,\ref{fig:Cernoxresistance} (b)). Nevertheless, the minor differences in the resistance behavior of different chips depicted in Fig.\,\ref{fig:Cernoxresistance} (b) requires a calibration of each chip for each pressure run, as will be described below.	All in all, our results show the Cernox chips can be used as temperature sensors in pressure experiments with high reliability and reproducibility.

\section{AC Specific heat: Theoretical background and Measurement Protocol}
\label{sec:ACtheory}

In the following, we want to focus on one possible application for which it is essential to determine temperatures with high sensitivity inside the pressure cell, namely when performing measurements of the specific heat of solids under pressure. As we employ here the method of AC calorimetry, this section will provide theoretical background information \cite{Sullivan68,Eichler79} on the AC calorimetric technique which is essential for understanding our measurement protocol.

\begin{figure}
				\includegraphics[width=0.9\textwidth]{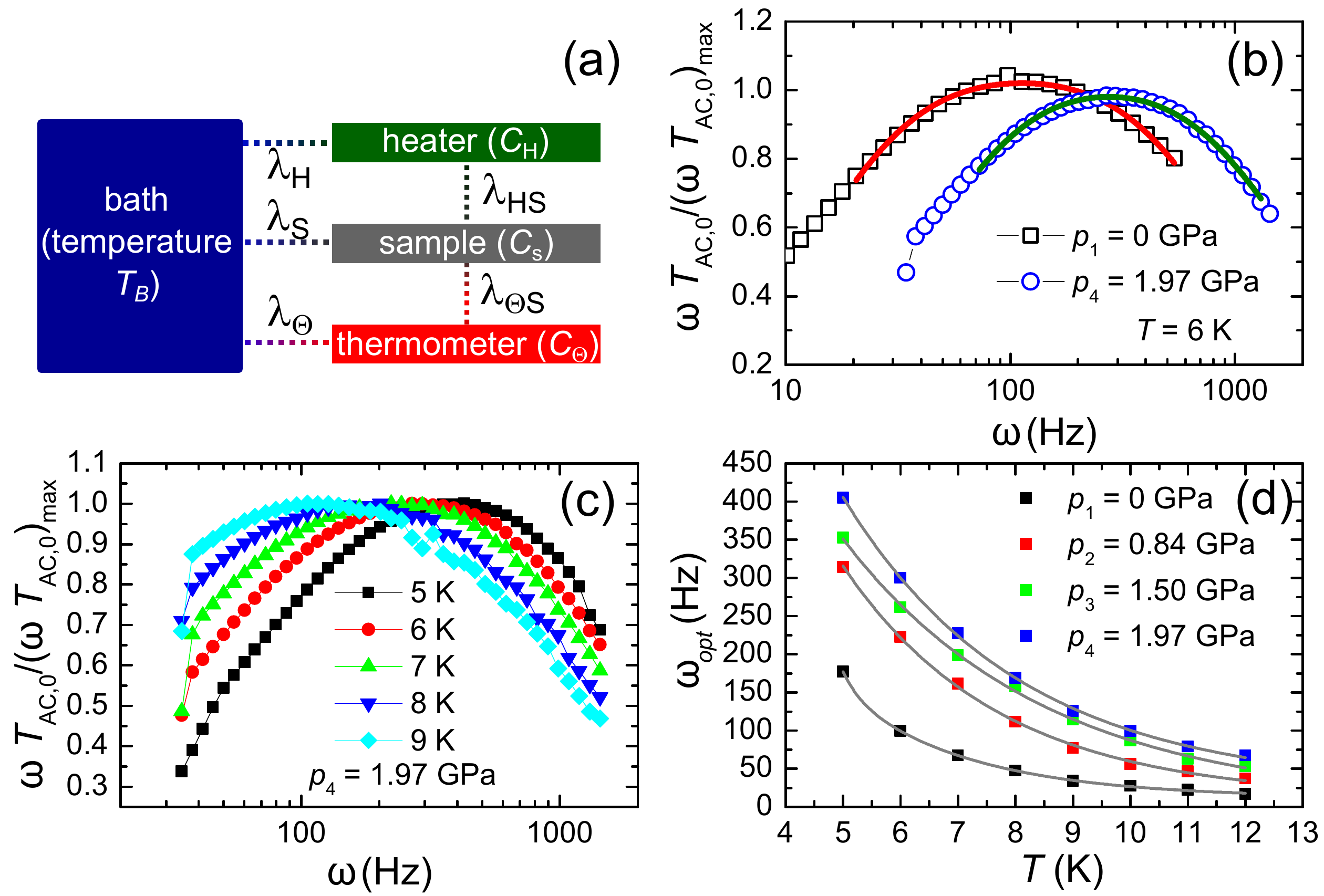} 
				\caption{(a) Schematic diagram illustrating heat flows from the sample to the heater (governed by the thermal conductivity $\lambda_{HS}$), from the sample to the thermometer ($\lambda_{\Theta S}$), and from sample, heater and thermometer to the bath ($\lambda_S$, $\lambda_H$ and $\lambda_\Theta$); (b) Normalized frequency response, i.e., the product of frequency $\omega$ and oscillation amplitude $T_{AC,0}$ normalized to its maximum value $(\omega T_{AC,0})_{max}$ vs. $\omega$, for a Pb sample at $T\,=\,6$\,K and $p\,=\,0\,$GPa and 1.97\,GPa; (c) Normalized frequency response for a Pb sample at different temperatures between 5\,K and 9\,K at $p\,=\,1.97\,$GPa; (d) Evolution of the optimal measurement frequency $\omega_{opt}$ as a function of temperature and pressure, obtained from the data presented in (b) and (c) (for details, see main text). Grey lines represent exponential fits to the $\omega_{opt}$ vs. $p$ data sets.}
				\label{fig:frequencydependence}
				\end{figure}

To extract absolute specific heat values from an AC calorimetry experiment, an understanding of the heat flow from the heater to the various components in the system is needed. On one hand, the heat is transferred from the heater (specific heat $C_H$) through the sample ($C_S$) to the thermometer ($C_{\Theta}$) which are connected via finite thermal conductances. In the following, the thermal conductances between sample and heater as well as thermometer and sample which govern the heat transfer are denoted as $\lambda_{HS}$ and $\lambda_{\Theta S}$, respectively. On the other hand, each component is also coupled to a bath with temperature $T_B$ (which corresponds to the temperature on the outside of the pressure cell) by finite thermal conductances, denoted by $\lambda_H$, $\lambda_S$ and $\lambda_\Theta$. A block diagram of this arrangement is shown in Fig.\,\ref{fig:frequencydependence} (a). Note that the couplings $\lambda_H$, $\lambda_S$ and $\lambda_{\Theta}$ are non-negligible in the present case, as the sample has to be embedded in a pressure medium inside the pressure cell for measurements under pressure. This, in general, means that the absolute values of the specific heat cannot be determined under pressure with high accuracy. Nonetheless, we will show below that the AC technique implemented in this work allows for a determination of specific heat value under pressure at a semi-quantitative level by choosing the right measurement frequency.

  When the heater is supplied with an AC voltage $U(t)\,=\,U_0 \sin(\frac{\omega}{2} t)$, it gives rise to an AC heating power $P(t)\,=\,P_0 \sin^2(\frac{\omega}{2}t)$ and the temperature of the sample will respond in the following manner:

		\begin{equation}
		T(t) \,=\, T_{DC} + T_{AC}(t).
		\end{equation}
		
$T_{DC}$ refers here to the time-independent increase of the sample temperature with respect to the bath, which is determined by the heating power as well as the coupling to the bath via $T_{DC}\,=\,T_B+\frac{P_0}{2\lambda_S}$. The second term $T_{AC}(t)\,=\,T_{AC,0} \sin(\omega t + \phi)$ describes the temperature oscillation of the sample which oscillates with twice the driving frequency of the heater. The amplitude of this oscillation $T_{AC,0}$ contains the information about the specific heat of the sample $C_S$. The sensitivity of an AC specific heat setup is particularly high for small samples (i.e. with small $C_S$), as $T_{AC,0}$ is inversely proportional to $C_S$. In detail, the relation of $T_{AC,0}$ to $C_S$ for a realistic model with finite thermal conductances was discussed in the works of Sullivan and Seidel (Ref. \citen{Sullivan68}), as well as Eichler (Ref. \citen{Eichler79}), and reads as

	\begin{eqnarray}
	T_{AC,0} \,&=&\, \frac{P_0}{2\omega C}\,\cdot\,F(\omega) \label{eq:frequency-response} \\
	\textnormal{with\ } F(\omega)\,&=&\,[1+2\,\cdot\,(\frac{\lambda_H}{\lambda_{HS}}+\frac{\lambda_\Theta}{\lambda_{\Theta S}})+\frac{1}{(\omega \tau_1)^2}+(\omega \tau_2)^2]^{-1/2} \label{eq:frequency-response2}\\
	\textnormal{and\ } C &=& C_H + C_\Theta + C_S, \tau_1=\frac{C}{\lambda_S}, \\
	\tau_2 &=& \sqrt{\tau_H^2+\tau_\Theta^2} \textnormal{\ with \ } \tau_H\,=\,C_H/\lambda_{HS} \textnormal{\ and \ } \tau_\Theta = C_\Theta/\lambda_{\Theta S}.
	\end{eqnarray}
	
	Thus, whenever the measurement frequency is choosen such that $(\omega \tau_1)^2\,\gg\,1$ and $(\omega \tau_2)^2\,\ll\,1$, with $\tau_1$ and $\tau_2$ corresponding to the thermal relaxation times to the bath and within the assembly of heater, sample and thermometer, respectively, then $F(\omega)\,\approx\,[1+2\,\cdot\,(\frac{\lambda_H}{\lambda_{HS}}+\frac{\lambda_\Theta}{\lambda_{\Theta S}})]^{-1/2}$. The error in determination of absolute values then is mainly determined by the ratios $\frac{\lambda_H}{\lambda_{HS}}$ and $\frac{\lambda_\Theta}{\lambda_{\Theta S}}$. However, sample, heater and thermometer are in intimate contact, whereas the heat path to the bath (i.e., the outside of the pressure cell) is long. This implies due to the geometrical arrangement that, to a first approximation, $\lambda_H\,\ll\,\lambda_{HS}$ and $\lambda_\Theta\,\ll\,\lambda_{\Theta S}$ (and we will show below that this assumption is verified in our setup), and therefore $F(\omega)\,\approx\,1$. It then follows that $T_{AC,0}\,=\,\frac{P_0}{2\omega C}$. 
	
	The frequency which meets these criteria is called the optimal measurement frequency $\omega_{opt}$. As $\tau_1$ and $\tau_2$ depend on the specific heat of the sample, as well as on thermal conductances $\lambda_{HS}, \lambda_S$ and $\lambda_{\Theta S}$, $\omega_{opt}$ will in general be a function of temperature and pressure, and will differ from sample assembly to sample assembly. Correspondingly, $\omega_{opt}$ has to be determined experimentally for each sample, temperature and pressure individually, prior to each measurement of the specific heat. It can be shown that $\omega_{opt}$ is the frequency at which $F(\omega)$ is maximal. As suggested by eq. \ref{eq:frequency-response}, direct experimental access to $F(\omega)$ is provided by measuring the frequency dependence of the quantity $\omega\,\cdot\,T_{AC,0}$ (called frequency response hereafter). In Figs. \ref{fig:frequencydependence} (b) and (c) we show examples of the frequency responses, normalized to their respective maximum, recorded with our setup when measuring the specific heat of elemental Pb (The specific heat results on Pb will be discussed in Sec.\,\ref{sec:specheatresults} in more detail.). First, we compare in Fig.\,\ref{fig:frequencydependence}\,(b) the frequency response, taken at $T\,=\,6$\,K at two different pressures ($p\,=\,0\,$GPa and 1.97\,GPa). Each frequency response (normalized to its maximum value) reveals a broad maximum at $\approx\,100\,$Hz and 300\,Hz, respectively, which we assign to the optimal measurement frequencies $\omega_{opt}$. Note that a broad maximum (or even a wide plateau) suggests that $(\omega \tau_1)^2\,\gg\,1$ and $(\omega \tau_2)^2\,\ll\,1$ across a wide frequency range which minimizes errors in the determination of absolute values of the specific heat. The observation of a broad maximum is therefore crucial for the determination of absolute values of the specific heat under pressure on a semi-quantitative level, as achieved with our setup. At the same time, at a fixed pressure, as shown in Fig.\,\ref{fig:frequencydependence}\,(c) for $p\,=\,$1.97\,GPa, we find that the broad maximum in the frequency response and thereby $\omega_{opt}$ shifts to lower frequencies with increasing temperature. The evolution of $\omega_{opt}$, determined from the numerical derivation of the frequency response data, with $T$ and $p$ is summarized in Fig.\,\ref{fig:frequencydependence}\,(d).
	
	The knowledge of the frequency response allows us to extract the relaxation times $\tau_1$ and $\tau_2$ of this particular assembly at different pressures. The solid lines in Fig.\,\ref{fig:frequencydependence}\,(b) show a fit of eq.\,\ref{eq:frequency-response2} to our experimental data, taken at 6\,K. The fits, which were performed with keeping $(\frac{\lambda_H}{\lambda_{HS}}+\frac{\lambda_\Theta}{\lambda_{\Theta S}})$ fixed to 0, are in very good agreement with our experimental data set. They yield $\tau_1\,=\,(0.047\,\pm\,0.001)\,$s and $\tau_2\,=\,(0.0016\,\pm\,0.0001)\,$s at ambient pressure and $\tau_1\,=\,(0.015\,\pm\,0.001)\,$s and $\tau_2\,=\,(0.0009\,\pm\,0.0001)\,$s at $p\,=\,1.97\,$GPa. Thus, the optimal measurement frequencies $\omega_{opt}$ fulfill the criteria mentioned above as $(\omega_{opt} \tau_1)^2\,\approx\,22\,\gg\,1$ and $(\omega_{opt} \tau_2)^2\,\approx\,0.02\,\ll\,1$ at $p\,=\,0\,$GPa and $(\omega_{opt} \tau_1)^2\,\approx\,20\,\gg\,1$ and $(\omega_{opt} \tau_2)^2\,\approx\,0.07\,\ll\,1$ at $p\,=\,1.97\,$GPa.	Our fitting results indicate that both relaxation times decrease with increasing pressure. This tendency is naturally expected, as the coupling to the bath, but also the coupling within the assembly likely increase under compression. Therefore, the increased optimal frequency $\omega_{opt}$ with applied $p$ is directly a consequence of the decreased relaxation times. The temperature dependence of $\omega_{opt}$ is less intuitive to understand as it depends on the temperature-dependent changes of specific heat as well as thermal conductivity of sample, heater as well as thermometer.
	
	It should be noted that the error in the determination of the absolute value of the specific heat can be estimated from the knowledge of $\tau_1$ and $\tau_2$. Equations \ref{eq:frequency-response} and \ref{eq:frequency-response2} suggest that any finite $\tau_1$ as well as any non-zero $\tau_2$ will give rise to an overestimation of the specific heat value by the factor $[1+\frac{1}{(\omega \tau_1)^2}+(\omega \tau_2)^2]^{1/2}$, if $\frac{\lambda_H}{\lambda_{HS}}+\frac{\lambda_\Theta}{\lambda_{\Theta S}}\,\approx\,$0. Our results of $\tau_1$ and $\tau_2$ correspond to an overestimate of specific heat by $\approx\,$3\,\% at ambient pressure, and $\approx\,$5\,\% at 1.97 GPa by our setup. This estimate shows that our setup can, in principle, deliver absolute values on a semi-quantitative level, despite a non-negligible coupling to the bath. Importantly, the analysis performed here shows that the overestimation of specific heat does not significantly change with pressure. This allows for the determination of changes (especially relative changes) of specific heat under pressure with higher accuracy. We confirm these conclusions from the analysis of the frequency response in Sec. \ref{sec:specheatresults}, where we present specific heat under pressure data on three different test cases and compare with ambient-pressure literature data taken under adiabatic conditions. 
	
	The theoretical background information given here explains the measurement protocol which we follow to determine the specific heat of a sample using the AC technique. It includes in total three separate, sequential temperature sweeps. First, we need to calibrate the Cernox thermometers inside the pressure cell at a specific pressure to quantify the DC temperature increase $T_{DC}$ of the sample which results from applying heat to the heater inside the cell. To this end, we place a calibrated thermometer outside on the pressure cell and record the resistance of the Cernox thermometer inside the pressure cell upon slow warming with a rate of $\approx\,0.25\,$K/min without any voltage applied to the heater inside the cell (see Sec.\,\ref{sec:Cernoxresistance}). In the second temperature sweep, we record the frequency response $\omega\,\cdot\,T_{AC,0}$ vs. $\omega$ as a function of temperature for the same, specific pressure. From this data, we extract $\omega_{opt}$ as a function of $T$ and typically fit this smooth data set with an exponential function $\omega_{opt}\,=\,\omega_0+A\exp(-T/t_1)$ with free parameters $\omega_0$, $A$ and $t_1$ (see grey lines in Fig.\,\ref{fig:frequencydependence} (d)). Within our measurement program, we adjust the measurement frequency continuously with temperature according to this exponential function during the third temperature sweep for a specific pressure. This ensures that the AC temperature oscillation $T_{AC,0}$ as a function of $T$ is always measured at the optimal measurement frequency which then allows us to infer the specific heat on a semi-quantitative level. 
		
\section{Specific heat under pressure: Results}
\label{sec:specheatresults}

		In the following, we demonstrate the wide applicability of the Cernox thermometers in measurements of specific heat under pressure by examining three test cases with very different transition temperatures, ranging from $T\,\approx\,7\,$K (superconducting transition in Pb) up to $T\,\approx\,130\,$K (magnetostructural transition in BaFe$_2$As$_2$), as well as very different amounts of entropy change.

		\subsubsection{Superconducting phase transition in elemental Pb}
		
		The first sample for a study of specific heat under pressure chosen here is elemental lead (Pb) which undergoes an ambient-pressure superconducting transition at a critial temperature $T_c\,=\,7.2\,$K. The shift of $T_c$ with pressure is well characterized in literature \cite{Bireckoven88} and therefore often utilized as a manometer at low temperatures. 
		
				\begin{figure}
				\includegraphics[width=0.8\textwidth]{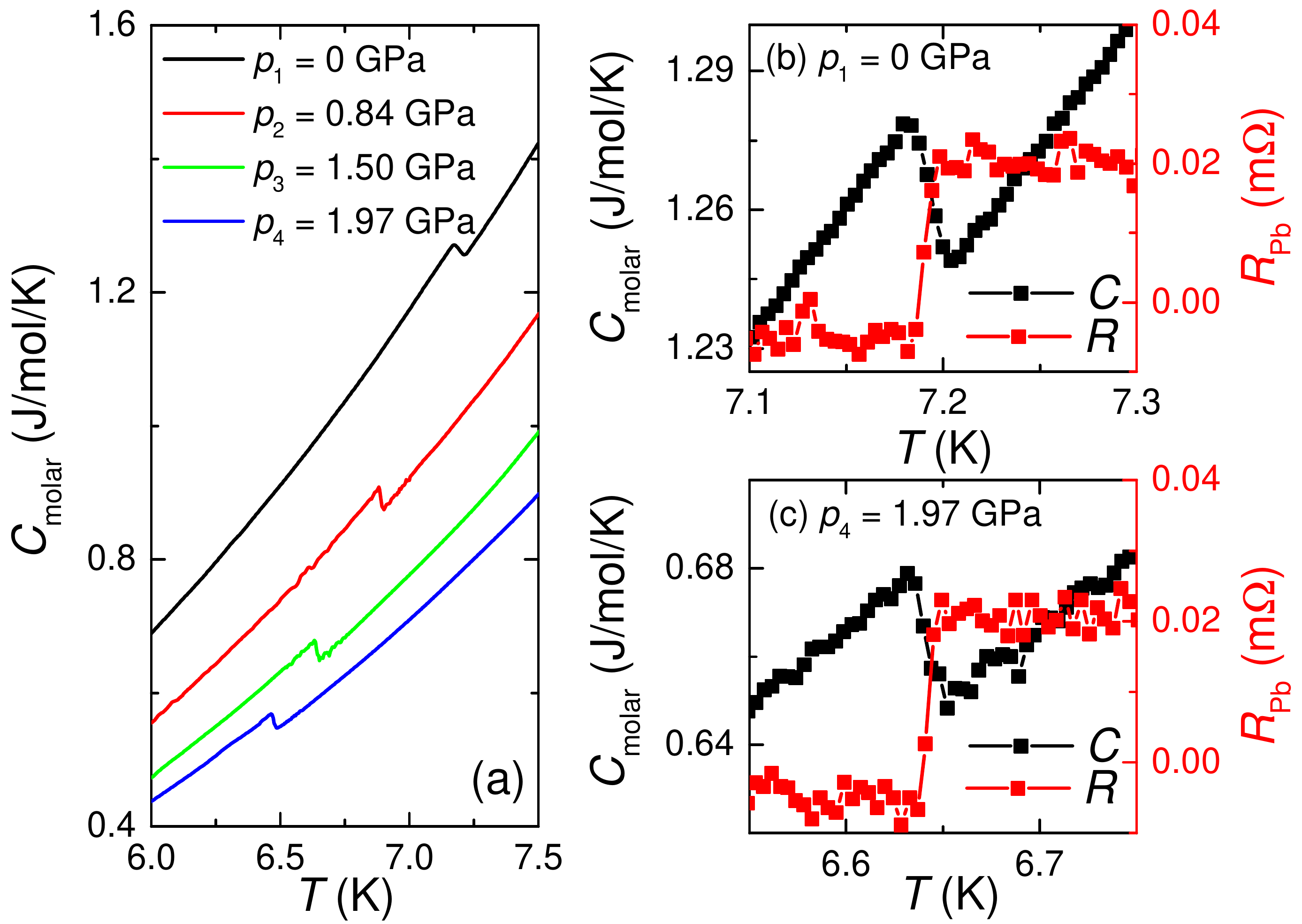} 
				\caption{(a) Molar specific heat $C_{molar}$ of elemental Pb as a function of temperature $T$ at four different pressures up to 1.97\,GPa; (b,c) Comparison of $C_{molar}(T)$ of Pb and the resistance $R_{Pb}(T)$ of the Pb pressure sensor at a pressure of 0 GPa (b) and 1.97\,GPa (c).}
				\label{fig:Pb}
				\end{figure}
				
		Figure \ref{fig:Pb} (a) shows our results of the specific heat, $C_{molar}$, of Pb at different pressures up to 1.97\,GPa. At all pressures, we find a jump-like change of $C_{molar}$ at a critical temperature $T_c$. This feature is characteristic for the mean-field type phase transition into the superconducting state in BCS superconductors. The critical temperature, extracted from our $C_{molar}$ data, is suppressed with increasing $p$, consistent with literature results \cite{Bireckoven88}. In addition, the overall specific heat is reduced upon pressurization, likely due to a combination of changes in the electronic density of states as well as lattice stiffening (see below for more details).
		
		To demonstrate the high accuracy in the determination of phase transition temperatures from our specific heat data, we compare in Figs. \ref{fig:Pb} (b) and (c) the specific heat of the Pb sample, placed between heater and thermometer, with the resistance of the Pb manometer, $R_{Pb}$, at lowest pressure ($p_1\,=\,0\,$GPa) and highest pressure ($p_4\,=\,1.97\,$GPa) of our experiment. At both pressures, the midpoint of the jump in $C_{molar}$ occurs at the same temperature at which the resistance clearly shows a jump-like change into the superconducting state. This also demonstrates that there are no significant pressure gradients in our pressure cell.
		
\begin{figure}
				\includegraphics[width=0.8\textwidth]{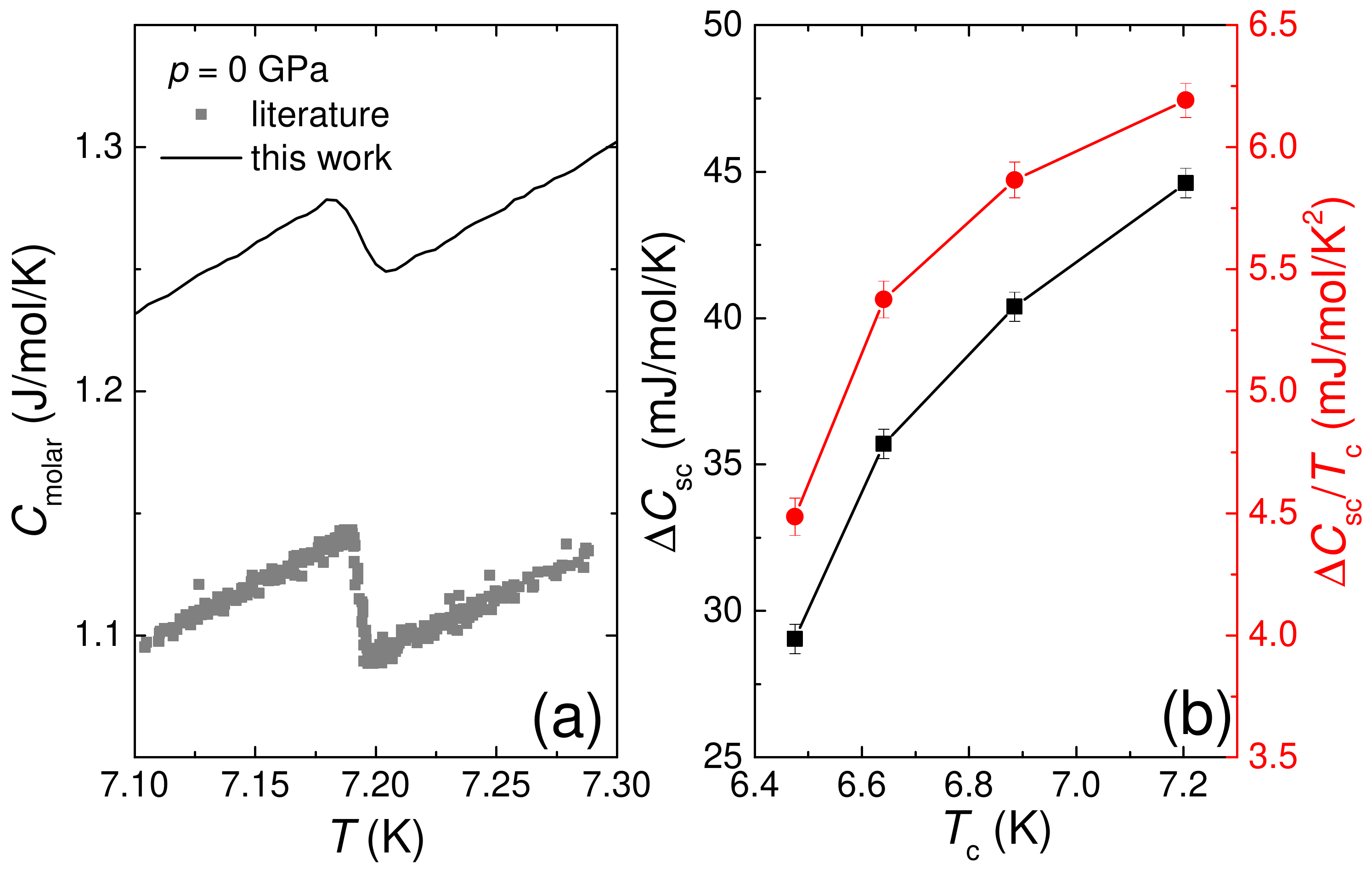} 
				\caption{(a) Comparison of ambient-pressure literature data \cite{Shiffman63} on the specific heat $C_{molar}$ of elemental Pb (grey symbols) and the specific heat obtained in the present work in the pressure cell at ambient pressure (black solid line); (b) Evolution of the superconducting jump size in the specific heat $\Delta C_{sc}$, as well as the jump size normalized to the critical temperature, $\Delta C_{sc}/T_c$, with $T_c$.}
				\label{fig:Pb-analysis}
				\end{figure}

		Next, we want to discuss to which extent our setup delivers a semi-quantitative determination of the specific heat of solids by comparing our results to literature results \cite{Shiffman63} on Pb (see Fig. \ref{fig:Pb-analysis}). Our data overestimates the absolute specific heat value by $\approx\,12\,\%$, compared to the literature results from Ref. \citen{Shiffman63}. As outlined in Sec.\,\ref{sec:ACtheory}, an overestimate of absolute specific heat values determined with the AC technique is a consequence of finite relaxation times $\tau_1$ and $\tau_2$. The overestimation factor of about 3\,\%, estimated from an analysis of the frequency responses in Sec. \,\ref{sec:ACtheory}, is of similar size as the overestimation found here from the comparison with literature data on Pb. Note that we did not correct our data for the specific heat of the addenda, i.e., of thermometer, heater and the tiny layers of glue. These additional contributions to the measured specific heat, which can be as large as 50\,\% of the total measured specific heat depending on the specific sample, its size, mass and shape (estimated by measuring the size of the addenda at ambient pressure using the relaxation-time method), likely give rise to the slightly larger overestimation of 12\,\% found empirically here. Clearly, despite the strong coupling to the bath due to the pressure medium and uncertainties in the size of the addenda, our data resembles literature results on a semi-quantitative level, i.e., within less than approximately a factor of 2. This upper limit was estimated empirically in the study of the three different test cases presented in this manuscript. The superconducting jump size extracted from our data at ambient pressure amounts to $\Delta C_{sc}\,\approx\,(44.6\,\pm\,0.5)\,$mJ/mol/K \cite{Shiffman63,Clement52}. This value is slightly smaller than reported values in literature ($\Delta C_{sc}\,=\,$52.9\,mJ/mol/K up to 57.5\,mJ/mol/K), but nevertheless matches on the same semi-quantitative level. 
		
		Upon pressurization, we find a significant reduction of $\Delta C_{sc}$ with increasing $p$ down to $\approx\,(29.0\,\pm\,0.5)\,$mJ/mol/K at 1.97\,GPa (see Fig.\,\ref{fig:Pb-analysis} (b)). For superconductors, the change in $\Delta C_{sc}$ is related on the one hand to a change in $T_c$ as well as to a change in the density of states at the Fermi level $N(E_F)$ \cite{Carbotte90}. Figure \ref{fig:Pb-analysis} (b) also includes a plot of $\Delta C_{sc}/T_c$ as a function of $T_c$. The strong change of $\Delta C_{sc}/T_c$ with $p$ by $\approx\,-\,25\,\%$ indicates that most of the change of $\Delta C_{sc}$ with $p$ can be attributed to changes of $N(E_F)$ with $p$, rather than to changes of $T_c$. Unfortunately, no literature data on the change of $N(E_F)$ in Pb with $p$ is available. Also the determination of the change of $N(E_F)$ under $p$ by extracting the Sommerfeld coefficient $\gamma$ from our specific heat data turns out to be not reliable due to the relatively high $T_c$ of Pb combined with a relatively low Debye temperature $\Theta_D\,\approx\,100\,$K at ambient pressure. Thus, we performed density-functional theory (DFT) calculations \cite{Hohenberg64,Kohn65} of the band structure of Pb up to 2\,GPa using PBEsol as exchange-correlation functional with spin-orbit coupling (SOC) effect as implemented in VASP \cite{Kresse96,Kresse96b}. At zero pressure, the theoretical lattice constant of 4.934\,\AA\ agrees very well with the experimental values of 4.95\,\AA. At 2 GPa, the lattice constant is reduced to 4.872\,\AA. We find a decrease of $N(E_F)$ from 0.5303\,states/eV/cell to 0.5121 \,states/eV/cell by 2\,GPa which corresponds to $\approx\,-\,3.5$\,\%. Even though this value is smaller than the one inferred from our specific heat measurements, both results are consistent in inferring a decrease of $N(E_F)$ with pressure. This supports our conclusion that in case of Pb changes of the specific heat anomaly $\Delta C_{sc}$ under $p$ result from a decrease of $N(E_F)$ with pressure.

		\subsubsection{Antiferromagnetic transition in the rare-earth compound GdNiGe$_3$}
		
		The specific heat anomaly at the superconducting transition in Pb as well as the transition temperature respond strongly to application of external pressure. In addition, the amount of entropy change is relatively small. For the next system, we chose GdNiGe$_3$ anticipating a weak response to pressure and a large change in entropy ($\approx\,R \ln(8)$) \cite{Mun10}. This allows us to show that changes of the absolute values of specific heat under pressure are not a result of an artifact due to modified coupling to the bath or within the assembly of sample, heater and thermometer, i.e., due to changing $\tau_1$ and $\tau_2$. This in turn allows us to establish a high accuracy in the determination of \textit{changes} of the specific heat under pressure. At the same time, a system with a phase transition occurring at higher transition temperature compared to the superconducting transition in elemental Pb is desired to prove high sensitivity of our setup at even higher temperatures.		
		
\begin{figure}
				\includegraphics[width=0.8\textwidth]{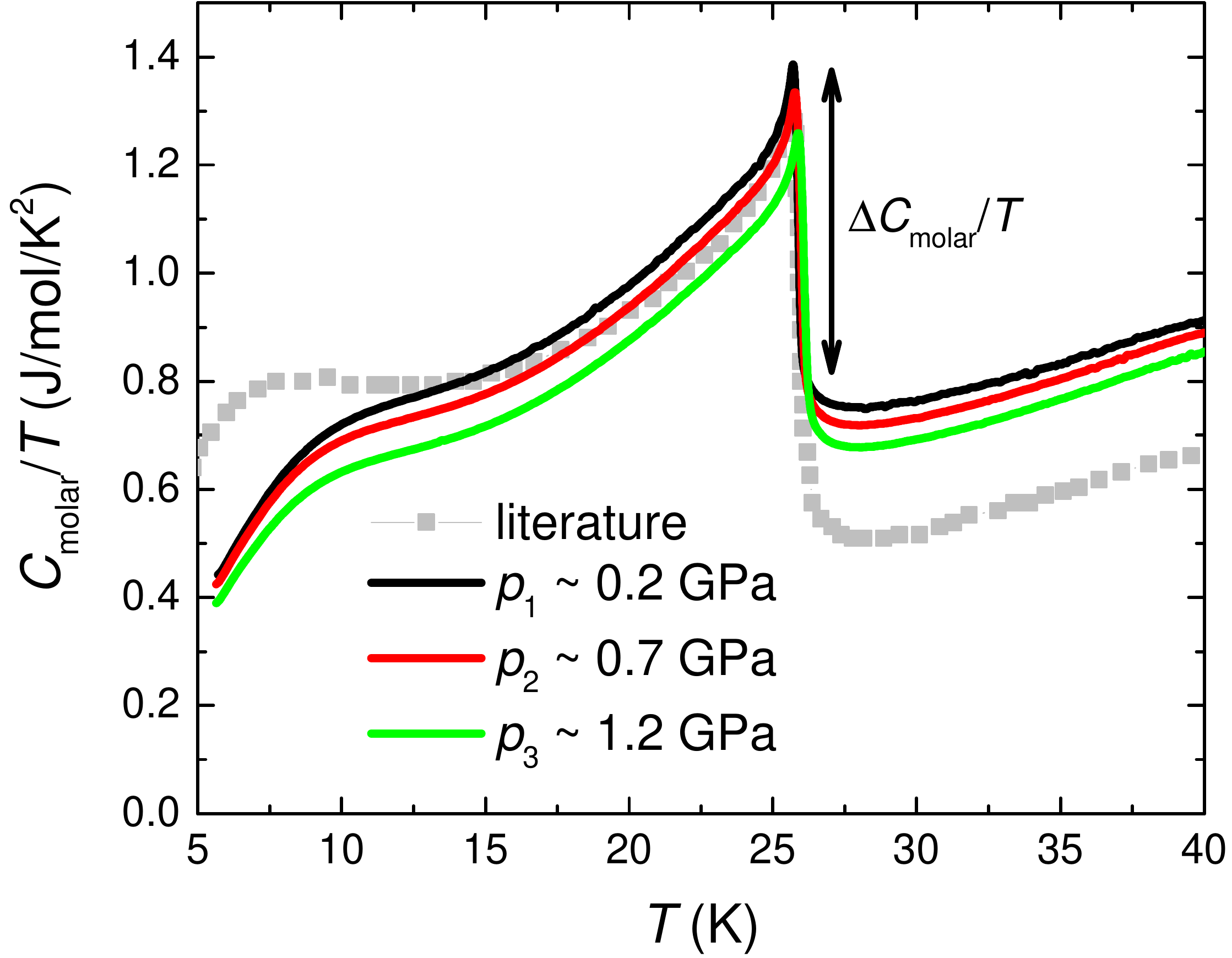} 
				\caption{Molar specific heat divided by temperature $C_{molar}/T$ as a function of temperature $T$ of GdNiGe$_3$ at different pressures up to 1.2 GPa. For comparison, literature data at ambient pressure \cite{Mun10} on this compound are shown in grey. The arrow indicates exemplary the jump size at the antiferromagnetic transition, $\Delta C_{molar}/T$, at lowest pressure $p\,\sim\,0.2$\,GPa (for a definition of criterion and evolution with pressure, see main text). }
				\label{fig:GdNiGe3}
				\end{figure}

		The rare-earth based GdNiGe$_3$ system has a single antiferromagnetic (afm) transition\cite{Mun10} at $T_N\,\approx\,$26\,K. Importantly, as the moment-carrying Gd is trivalent in this compound and therefore has a Hund's rule $J\,=\,S\,=\,7/2$ ground state ($L\,=\,0$), the compound lacks any magneto-crystalline anisotropy or splitting of the Hund's rule ground state multiplet. Experiments \cite{Mun10} confirmed that GdNiGe$_3$ shows an almost isotropic susceptibility in the paramagnetic state with an effective moment $\mu_{eff}\,=\,8.0\,\mu_B$/Gd$^{3+}$, close to the free-ion value of 7.94\,$\mu_B$/Gd$^{3+}$. Correspondingly, specific heat measurements \cite{Mun10} showed a single $\lambda$-shaped peak at $T_N$ (see grey symbols in Fig.\,\ref{fig:GdNiGe3} for a reproduction of these data). The magnetic entropy, $S$, extracted from measurements of the specific heat was found to be almost constant at $T\,>\,T_N$ with $S\,=\,$17\,J/(mol$\cdot$K), i.e., close to the expected value of $R \ln(8)$. This result is fully consistent with the absence of crystal-field effects in this compound. As the magnetism of this compound can be well understood in terms of localized 4f moments which interact via RKKY (Ruderman-Kittel-Kasuya-Yosida) interaction \cite{Ruderman54}, the response to hydrostatic pressure is expected to be relatively weak. This, together with the well-defined entropy in the paramagnetic state and a high transition temperature, makes this system a suitable reference system for a study of the specific heat under pressure. 	
		
		The results of our specific heat study on GdNiGe$_3$ under pressure are shown in Fig.\,\ref{fig:GdNiGe3} in a $C_{molar}/T$ vs. $T$ representation. For comparison, we included the literature specific heat data on this compound at ambient pressure, taken from Ref. \citen{Mun10}. At all measured pressures, our data nicely reveal the $\lambda$-shaped phase transition anomaly at $T\,\approx\,26\,$K. In addition, we also find a small hump in the specific heat below 10\,K. Such a hump in the specific heat at temperatures well below the ordering temperatures was found in various Gd-based systems \cite{Kong14} and was explained by modelling the specific heat of a $(2J+1)$ multiplet in a mean-field approach \cite{Bouvier91,Blanco91}. The comparison of our specific heat data with literature in terms of absolute values indicates an $\approx\,$10\% to 40\% overestimation of the specific heat for $T\,>\,15\,$K. For $T\,<\,15\,$K, we find an underestimate of $C_{molar}/T$. The reason for this behavior is unclear at present, as eqs.\,\ref{eq:frequency-response} and \ref{eq:frequency-response2} do not allow an underestimate. However, this additional data set on GdNiGe$_3$ also confirms that we are not only highly sensitive in tracing phase transitions even at higher temperatures, but also that we can determine absolute specific heat values within less than a factor of 2 deviation from literature.
				
				\begin{figure}
				\includegraphics[width=0.8\textwidth]{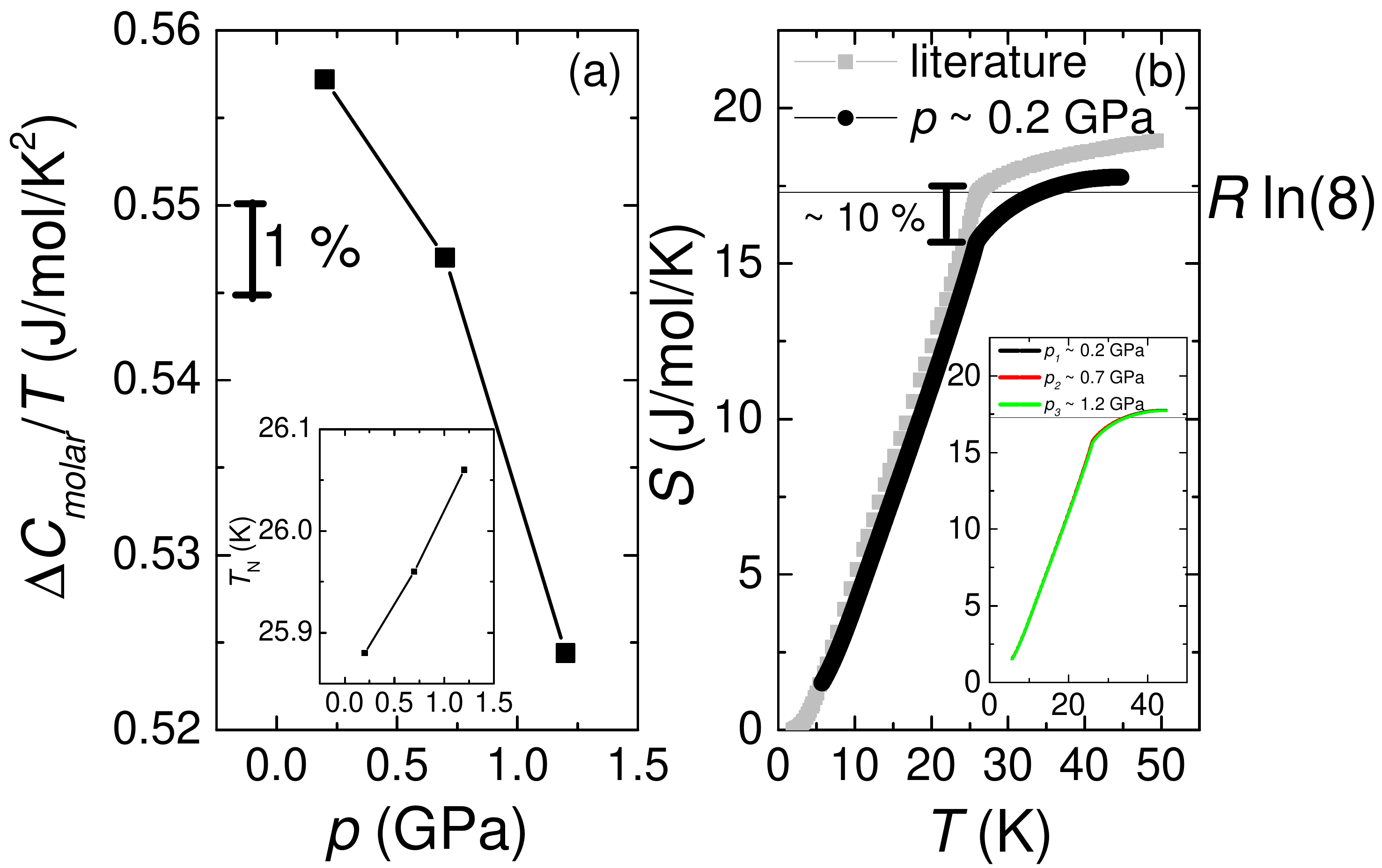} 
				\caption{(a) Change of jump size in specific heat at $T_N$, $\Delta C_{molar}/T$, and $T_N$ (inset) with pressure; (b) Comparison of estimated magnetic entropy at 0.2 GPa (present experiment, black circles) with literature results\cite{Mun10} at ambient pressure (grey squares). Inset: Estimated magnetic entropy at different pressures up to 1.2\,GPa.}
				\label{fig:GdNiGe3-analysis}
				\end{figure}
				
		In the following, we focus on the relative evolution of the specific heat with pressure. The overall specific heat values at each temperature are reduced upon applying pressure. Similar to the case of Pb, we assign this reduction to changes of the lattice and electronic specific heat. More importantly, however, we find that the $\lambda$-shaped peak, as well as the low-temperature hump are almost unaffected by pressure. This relates to the position of the anomalies as well as the size and shape of the anomaly. To quantify this statement, we show in Fig.\,\ref{fig:GdNiGe3-analysis} (a) the evolution of the jump size of the $\lambda$-shaped anomaly, $\Delta C_{molar}/T$, as well as the transition temperature $T_N$ (inset) with pressure. Whereas $T_N$ is extracted from the minimum in the derivative of the $C_{molar}/T$ data, $\Delta C_{molar}/T$ is calculated by the difference of $C_{molar}/T$ values at those temperatures at which d($C_{molar}/T$)/d$T$\,=\,0.1\,(d($C_{molar}/T$)/d$T$)$_{T_N}$ (see arrow in Fig.\,\ref{fig:GdNiGe3}). We find a slight decrease of $\Delta C_{molar}/T$ with pressure by $\approx\,-\,$6\,\% and a small increase of $T_N$ by less than 1\,\% within a pressure range of $\approx\,$1\,GPa. We want to emphasize that the relative change in $\Delta C_{molar}/T$ is tiny compared to the specific heat changes observed under pressure in elemental Pb. At present, this tiny change of the specific heat features in GdNiGe$_3$ cannot unequivocally be assigned to a single origin: Either the change is indeed related to changes of physical properties under pressure (see Refs. \citen{Bouvier91,Blanco91} for theoretical discussions of specific heat features in Gd-based compounds on a mean-field level), or the change is an artifact arising from uncertainties in the absolute values determined with the AC calorimetric technique related to changes in the relaxation times $\tau_1$ and $\tau_2$ with pressure. Most likely, both factors actually play a role here, but more importantly none of them gives rise to changes in the specific heat beyond $\approx\,$6\,\% when changing pressure by 1\,GPa. Thus, we can conclude from our specific heat measurements on GdNiGe$_3$ that changes in the specific heat of more than $\approx\,$6\,\% within 1\,GPa, in particular at phase transitions, can reliably be attributed to changes of physical properties, rather than to instrumental artifacts. We also note that in principle the addenda contribution can change with pressure. This change is explicitly included in the error bar given above. However, in general it is reasonable to assume that in first approximation the specific heat of the addenda does not change with pressure.
		
		 To extract the magnetic entropy and changes of this quantity with pressure from the present data set, non-magnetic (phononic and electronic) contributions need to be subtracted. These contributions are typically obtained by measuring the specific heat of a non-magnetic reference sample if available. In this case, YNiGe$_3$ serves as a suitable non-magnetic reference system the ambient-pressure specific heat of which was reported in Ref. \citen{Mun10}. As an independent measurement of YNiGe$_3$ with our AC calorimetric setup would require a new assembly with different relaxation times $\tau_1$ and $\tau_2$ which likely give rise to different error in the determination of absolute specific heat values compared to our values on GdNiGe$_3$, this approach to determine the non-magnetic contributions is not suitable in the present case. However, assuming that the overestimation factor is, in first approximation, temperature-independent and contributions from heater and thermometer to our measured specific heat are comparably negligible, we can rescale the reported ambient-pressure data on YNiGe$_3$ such that it almost matches our specific heat data on GdNiGe$_3$ at $T\,\gg\,T_N$ and $p\,\sim\,0.2\,$GPa. This procedure allows us to provide an estimate on the magnetic entropy from our data set (see Fig.\,\ref{fig:GdNiGe3-analysis} (b)). Our estimate of the magnetic entropy yields $S\,\sim\,15.7\,$J/mol/K at $T\,=\,T_N$ which corresponds to 90\% of the expected $S\,=\,R\,\ln(8)$. Even if this analysis can only provide a rough estimate of the entropy due to the uncertainties involved in the determination of the non-magnetic contributions, it confirms that we can determine not only specific heat, but also entropies on the same semi-quantitative level. When now discussing changes of the entropy as a function of pressure, we have to make further assumptions on how the non-magnetic contributions are affected by pressure. The change of the non-magnetic contributions reveals itself e.g. in the measured specific heat at $T\,\gg\,T_N$ which indicates a sizable reduction of $C_{molar}/T$ with $p$. To account for this change, we make the reasonable assumption that changes in the Sommerfeld coefficient $\gamma$ as well as the Debye lattice constant $\beta$ give rise to changes in the non-magnetic specific heat via $C_{molar}/T\,\propto\,\gamma + \beta T^2$. We now apply a temperature-independent as well as a quadratic correction to the $C_{molar}/T$ data of YNiGe$_3$ such that it matches our specific heat data on GdNiGe$_3$ at $T\,\gg\,T_N$ for $p_2$ and $p_3$ individually and subtract the so-derived non-magnetic contributions from our experimental data on GdNiGe$_3$. These estimates of the magnetic entropy are shown in the inset of Fig.\,\ref{fig:GdNiGe3-analysis}. We do not find any significant changes of the estimated magnetic entropy with pressure. This result is consistent with the almost unchanged size of specific heat anomaly $\Delta C_{molar}/T$ with $p$ and provides further evidence that changes of specific heat and entropies with pressure can be estimated with comparably high accuracy.
		 		
		\subsubsection{Structural/magnetic transition in the iron-pnictide BaFe$_2$As$_2$}
		
		Finally, to further demonstrate the sensitivity of our setup at even higher temperatures (above 100\,K), we present specific heat measurements under pressure on BaFe$_2$As$_2$. This material undergoes a structural and antiferromagnetic transition at $T_{s,N}\,\approx\,130\,$K from a tetragonal-paramagnetic to an orthorhombic-antiferromagnetic state \cite{Rotter08}. In the BaFe$_2$As$_2$ system either chemical substitution (e.g. of Fe by Co) \cite{Ni08,Canfield10,Sefat08} or pressure \cite{Colombier09} suppress this structural-magnetic transition and unconventional superconductivity, with critical temperatures up to $\approx\,22\,$K, emerges.
		
				\begin{figure}
				\includegraphics[width=0.8\textwidth]{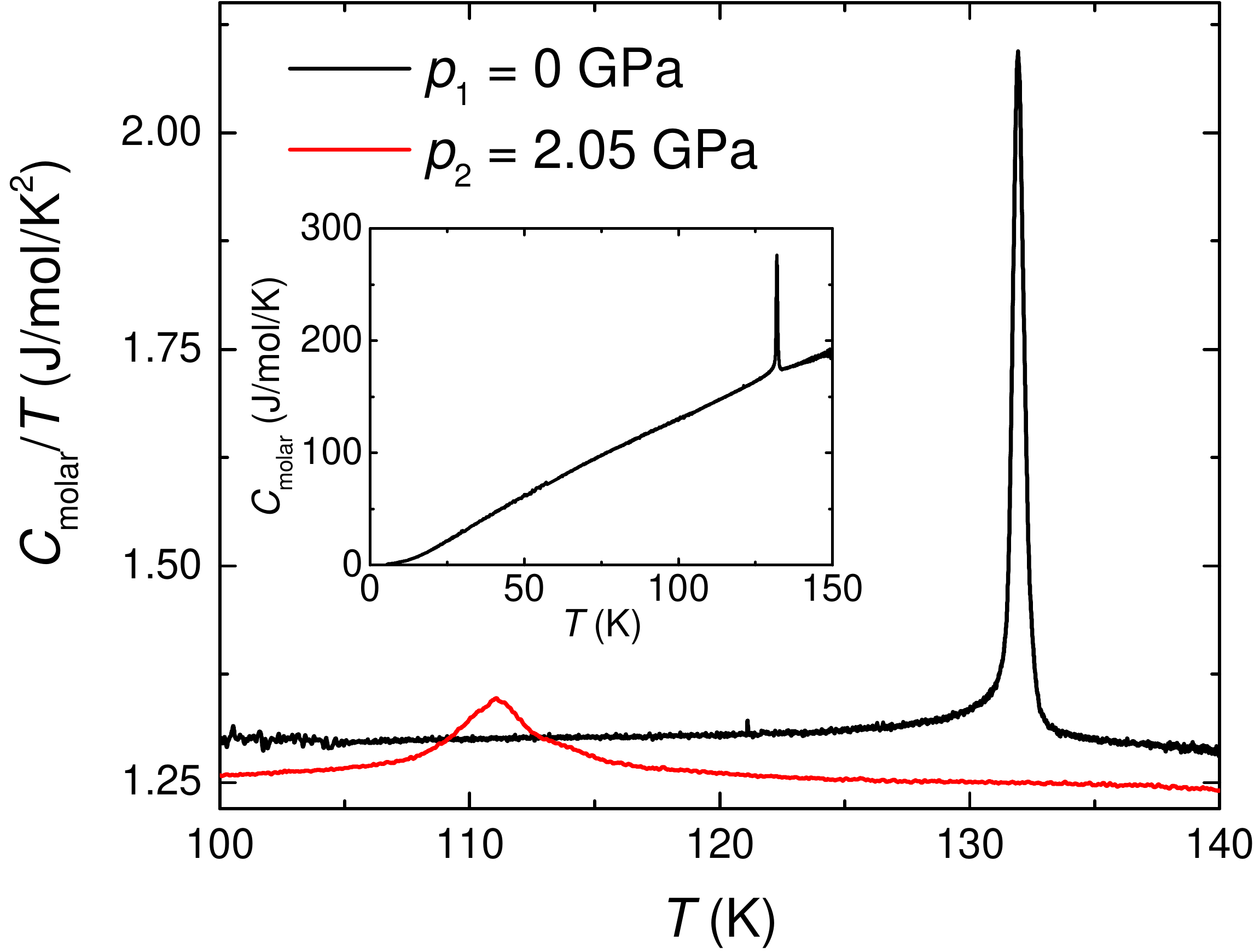} 
				\caption{Molar specific heat divided by temperature, $C_{molar}/T$, as a function of temperature $T$ of BaFe$_2$As$_2$ at ambient pressure as well as at $p\,=\,2.05$\,GPa; Inset: Specific heat, $C_{molar}$, of BaFe$_2$As$_2$ at ambient pressure on expanded scale.}
				\label{fig:Ba122}
				\end{figure}
				
		Figure \ref{fig:Ba122} shows our results for the specific heat of BaFe$_2$As$_2$ in the pressure cell at ambient pressure as well as at $p\,=\,$2.05\,GPa. Our ambient-pressure data show a very sharp peak at the structural-magnetic transition at $T\,\approx\,132\,$K. Even if the size of the phase transition is strongly reduced by the application of $p\,=\,2.05\,$GPa, indicating a strongly reduced entropy change at the phase transition with pressure, we can still clearly resolve the phase transition at a lower temperature $T\,\approx\,112\,$K. The decrease in the phase transition temperature by $\approx\,-10\,$K/GPa agrees very well with earlier reports of the pressure dependence of $T_{s,N}$ based on resistance data \cite{Colombier09}. More importantly, not only can we resolve the high-temperature anomaly, but we can also measure the specific heat across the entire temperature range 5\,K$\,\le\,T\,\le\,$150\,K by using a single thermometer (see inset of Fig.\,\ref{fig:Ba122}). This is a clear advantage of using the Cernox thermometers as temperature sensors for AC specific heat measurements rather than RuO$_2$ thermometers which are inherently sensitive only in a much more limited, low-temperature range. The setup presented here will therefore allow in the future to measure the specific heat under pressure of systems which show a cascade of phase transitions at very different temperatures, such as e.g. the superconducting as well as magnetic-structural transition in Co-doped BaFe$_2$As$_2$.
		
\section{Summary}

In conclusion, we studied the response of Cernox thermometers to external pressure in piston-pressure cells up to 2\,GPa. These thermometers are frequently used in low-temperature experiments due to their high sensitivity. We find that the sensitivity of the Cernox thermometers remains high under pressures up to 2\,GPa. In addition, they are mechanically robust and survive numerous pressure cycles. Thus, our study shows that these temperature sensors can be used to measure temperatures inside the pressure cell with high accuracy. As a possible application, for which this high sensitivity is essential, we present in detail the use of these thermometers in measuring the specific heat of solids under pressure. By studying three different test cases (elemental Pb, GdNiGe$_3$ and BaFe$_2$As$_2$), we show that the high sensitivity of the Cernox thermometers allows to measure specific heat of solids under pressure across a wide temperature range as well as wide range of entropy changes. Therefore, by using Cernox thermometers, it will be possible in the future to study systems which show a cascade of phase transition across a wide temperature range by specific heat under pressure in a piston-pressure cell, possibly even up to room temperature. In addition, we demonstrate that our setup does not only allow to trace phase transitions, but is also very accurate in determining changes of the specific heat as a function of pressure.

\begin{acknowledgments}
We thank E. Mun and N. Ni for growing the GdNiGe$_3$ and BaFe$_2$As$_2$ crystals used in the study. This work was carried out at Iowa State University and supported by Ames Laboratory, US DOE, under Contract No. DE-AC02-07CH11358. G.D.'s efforts were partially funded by the Gordon and Betty Moore Foundation's EPiQS Initiative through Grant No. GBMF4411. L.X. was supported, in part, by the W. M. Keck Foundation. 
\end{acknowledgments}

\bibliographystyle{modaps}
%\bibliography{Lit}

\end{document}